\documentclass[lettersize,journal]{IEEEtran}
\usepackage{amsmath,amsfonts}
\usepackage{algorithmic}
\usepackage{algorithm}
\usepackage{array}
\usepackage[caption=false,font=normalsize,labelfont=sf,textfont=sf]{subfig}
\usepackage{textcomp}
\usepackage{stfloats}
\usepackage{url}
\usepackage{verbatim}
\usepackage{graphicx}
\usepackage{cite}

\usepackage{booktabs}
\usepackage{pifont}
\usepackage{multirow}
\usepackage{amsmath}

\usepackage{threeparttable}

\begin{document}

\title{DEMUX: Boundary-Aware Multi-Scale Traffic Demixing for Multi-Tab Website Fingerprinting}

\author{Yali Yuan, Yaosheng Liu, Qianqi Niu, Guang Cheng
      
\thanks{Yali Yuan and Yaosheng Liu contributed equally to this work.} 
\thanks{Yali Yuan, Qianqi Niu and Guang Cheng are with the School of Cyber Science and Engineering, Southeast University, Nanjing 211189, China.}
\thanks{Yaosheng Liu is with the School of Artificial Intelligence, Southeast University, Nanjing 211189, China.}
\thanks{Guang Cheng is the corresponding author. (e-mail: gcheng@seu.edu.cn).}}

\maketitle

\begin{abstract}

Website fingerprinting (WF) attacks infer the websites visited by users from encrypted traffic in anonymous networks such as Tor. Existing deep learning methods achieve high accuracy under the single-tab assumption but degrade substantially when users open multiple tabs concurrently, producing interleaved traffic that transforms WF into an implicit demixing problem. We identify three structural requirements for effective multi-tab demixing, namely signal integrity at segment boundaries, multi-scale local modeling, and relative temporal association of dispersed fragments, and show that no prior method satisfies all three simultaneously. We propose DEMUX, a designed framework that addresses these requirements through three tightly coupled components. A Boundary Preserving Aggregation Module employs overlapping window partitioning with joint packet-level and burst-level feature extraction. A Multi-Scale Parallel CNN captures heterogeneous temporal patterns via parallel branches. A two-stage Transformer encoder with Rotary Positional Embedding enables robust cross-window fragment association. The Boundary Preserving Aggregation Module additionally serves as a plug-and-play preprocessor that consistently improves existing baselines without architectural modification. Extensive experiments across closed-world, open-world, defense-augmented, dynamic-tab, and cross-configuration settings demonstrate that DEMUX achieves state-of-the-art performance. In the challenging closed-world 5-tab setting, DEMUX attains a P@5 of 0.943 and MAP@5 of 0.961, outperforming the strongest baseline by 9.2 and 6.2 percentage points respectively, confirming its strong robustness in complex multi-tab demixing scenarios.
\end{abstract}


\section{Introduction}
\label{sec:introduction}

\begin{figure}[t]
    \centering \includegraphics[width=0.5\textwidth]{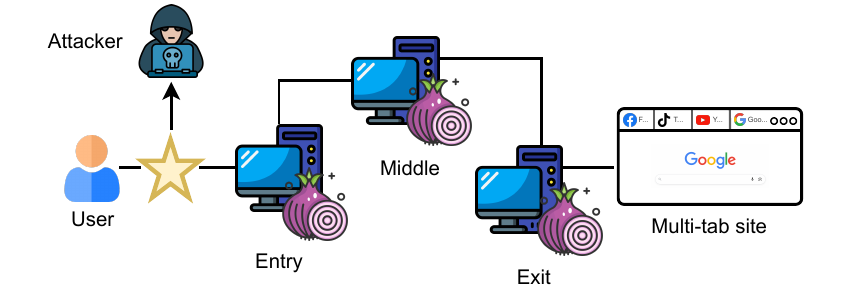}
    \caption{Website fingerprinting attackers monitor the encrypted traffic between the user and the Tor entry node to infer which websites the user is browsing.}
    \label{fig:tm}
\end{figure}

\IEEEPARstart{W}{ith} the rapid expansion of Internet technologies, network communication has become integral to daily life. However, the inherent openness of the Internet introduces significant privacy and security risks, driving the development and widespread adoption of anonymous communication systems such as The Onion Router (Tor)~\cite{dingledine2004tor}. Although Tor employs multi-layer encryption to protect user anonymity, it remains susceptible to the analysis of encrypted traffic patterns~\cite{xu2022seeing}. By exploiting observable features such as packet timing and size, an adversary can infer the websites a user visits, an attack known as Website Fingerprinting (WF)~\cite{panchenko2011website}, as illustrated in Fig.~\ref{fig:tm}. Over the past decade, WF attacks have advanced substantially, evolving from hand-crafted features with conventional classifiers~\cite{hayes2016k,panchenko2016website,oh2019p,de2020trafficsliver} to the deep learning paradigm established by Deep Fingerprinting (DF)~\cite{sirinam2018deep}, which feeds raw packet-direction sequences into stacked convolutional layers to learn discriminative representations in an end-to-end manner. Under the single-tab assumption, wherein each trace corresponds to exactly one target website, these approaches achieve identification rates exceeding 90\% in controlled settings.

Real-world browsing, however, rarely conforms to the single-tab assumption. Users routinely open multiple websites concurrently, producing a single observed trace that superimposes the traffic of $k$ overlapping websites, where $k$ is dynamic and unknown to the attackers. This transforms WF from a classification problem into a \emph{traffic demixing} problem, in which the attackers must recover the identities of all $k$ websites from a single entangled observation without prior knowledge of $k$ or the temporal boundaries between constituent flows. This shift is qualitative rather than merely quantitative. Rather than an easy scaling of the single-tab task, multi-tab demixing poses a fundamentally different problem whose structure imposes architectural constraints that no existing method was designed to satisfy.

To understand what these constraints entail, we systematically analyze the architectural gap between the multi-tab demixing problem and existing WF architectures. Prior work has largely built upon the DF architecture. This CNN-based backbone has underpinned a broad family of subsequent single-tab methods~\cite{rimmer2018automated,bhat2018var,cui2020more,wang2022snwf,ling2022towards,zou2025toward,li2025cross}. Although efforts to extend WF to multi-tab settings have leveraged attention mechanisms and structured traffic representations to model long-range dependencies in concurrent traffic~\cite{guan2021bapm,deng2023robust,jin2023transformer,deng2025towards}, these methods likewise retain the single-scale local feature extraction design of DF. Among all existing multi-tab methods, the ARES framework~\cite{deng2023robust,deng2025towards} has demonstrated the strongest overall performance. ARES'23~\cite{deng2023robust} established a Transformer-based multi-tab WF baseline and contributed the benchmark datasets that have since become standard in the field. ARES'25~\cite{deng2025towards} subsequently introduced an improved Transformer backbone that extracts both packet-level and burst-level features per window, achieving state-of-the-art results across multiple evaluation scenarios. However, even ARES shares foundational limitations inherited from the DF architecture. This analysis reveals three requirements that any effective demixing architecture need to satisfy simultaneously, yet no existing method fulfills all three. We summarize these requirements as follows.
\begin{enumerate}
    \item[\textbf{R1.}] \textbf{Signal integrity at segment boundaries.} In multi-tab traffic, burst-boundary transitions, where the dominant contributing website changes, carry the most discriminative cross-source switching signals. Fixed non-overlapping window segmentation, inherited from single-tab methods, systematically fragments these transitions across adjacent windows, destroying the very evidence required to distinguish co-occurring sources.
    \item[\textbf{R2.}] \textbf{Multi-scale local modeling.} Traffic fragments from different websites coexist at diverse temporal scales within a mixed trace. Burst patterns and periodic loading rhythms demand simultaneously short and long receptive fields, a diversity that single-scale CNN backbones cannot accommodate.
    \item[\textbf{R3.}] \textbf{Relative temporal association of dispersed fragments.} Fingerprint evidence for a single website may appear in fragments scattered throughout the traffic, with arbitrary concurrent traffic in between. Absolute positional encodings tie position indices to the superimposed mixture rather than to any individual source, making them structurally inadequate for cross-window fragment association under varying tab compositions.
\end{enumerate}

To address these requirements, we propose \textbf{DEMUX}, a multi-tab WF framework whose architecture is derived directly from the structural properties of the demixing problem. DEMUX departs from prior work in two principled ways. First, it replaces fixed non-overlapping segmentation with an overlapping window partitioning strategy that ensures boundary-adjacent burst transitions are always captured intact within at least one window, directly satisfying (R1). Second, it replaces the single-scale CNN backbone with a \emph{Multi-Scale Parallel CNN (MSP-CNN)} that simultaneously extracts fine-grained burst-level patterns and coarse-grained periodic structures, thereby addressing (R2). The resulting multi-scale representations are fused via pointwise convolution and processed by a \emph{Transformer encoder equipped with Rotary Positional Embedding (RoPE)}, which models long-range temporal dependencies through relative positional offsets rather than absolute indices, directly satisfying (R3).
Beyond its role within DEMUX, the overlapping window strategy is designed as a \emph{plug-and-play, model-independent} module, termed the Boundary Preserving Aggregation Module, that can replace the aggregation component of any WF architecture relying on temporal or burst-derived features. We validate its generality by integrating Boundary Preserving Aggregation Module into several representative baselines and demonstrate consistent and statistically significant improvements across all of them.
The contributions of this paper are summarized as follows.
\begin{itemize}
\item We propose DEMUX, a multi-tab WF framework co-designed to satisfy the three structural requirements (R1--R3) that multi-tab demixing imposes on any effective architecture, requirements that no existing method fulfills. DEMUX integrates three components, namely overlapping boundary-preserving aggregation, multi-scale parallel convolution, and a RoPE-enhanced Transformer encoder, that work in concert to achieve robust demixing under obfuscated concurrent traffic.

\item We propose a plug-and-play, universally applicable Boundary Preserving Aggregation Module that replaces fixed non-overlapping segmentation with overlapping window partitioning, preserving burst-boundary transition signals critical for demixing. While Boundary Preserving Aggregation Module serves as the core component of DEMUX for satisfying (R1), it is generalizable to any WF pipeline relying on temporal or burst-derived features. We validate this generality by integrating Boundary Preserving Aggregation Module into representative baselines including DF, TMWF, and ARES'25, achieving consistent improvements without any architectural modification.

\item We conduct evaluations across closed-world, open-world, defense-augmented, dynamic-tab, and cross-configuration settings. DEMUX achieves state-of-the-art performance across all settings. In the closed-world 5-tab setting, DEMUX attains a P@5 of 0.943 and MAP@5 of 0.961, outperforming the strongest baseline by 9.2 and 6.2 percentage points respectively. Under the most challenging TrafficSliver defense, DEMUX maintains a P@2 of 0.940, exceeding the next-best competitor by 2.5 points. Notably, these advantages widen as the number of concurrent tabs increases, confirming that the strong robustness of DEMUX in practical multi-tab demixing scenarios.
\end{itemize}

\section{Related Work}
Website fingerprinting has been extensively studied as a traffic analysis technique in the context of anonymity networks and encrypted communications. We organize existing work along the technical lineage most relevant to DEMUX, progressing from early statistical approaches, through classical machine learning methods, then the deep convolutional architecture introduced by DF and its subsequent extensions, to recent multi-tab methods.

\subsection{Single-Tab Website Fingerprinting}
\subsubsection*{Early Statistical and Classical Machine Learning Methods}
Early WF research paired hand-crafted features with conventional classifiers. Representative examples include the kNN-based attack~\cite{wang2014effective}, which evaluated WF under large open-world settings, and CUMUL~\cite{panchenko2016website}, which maps traces into a compact cumulative representation to enable scalable classification. Hayes and Danezis~\cite{hayes2016k} proposed k-fingerprinting, a random forest based approach designed for robustness and scalability, and evaluated its performance against several WF defenses. These works established foundational WF threat models and evaluation protocols, but uniformly assume that a traffic trace corresponds to a single website visit, an assumption that becomes untenable under real-world multi-tab browsing.

\subsubsection*{The DF Architecture and Its Variants}
A major advance occurred with Deep Fingerprinting (DF)~\cite{sirinam2018deep}, which showed that stacked 1D convolutional networks operating directly on raw packet direction sequences, represented as fixed-length $\pm1$ arrays of up to 5000 packets, can yield highly discriminative fingerprint representations without manual feature engineering. DF demonstrated strong performance on Tor traffic and was the first attack effective against WTF-PAD~\cite{juarez2015wtf}. This approach of treating the entire trace as a single 1D sequence and processing it with a single-scale convolutional backbone has since become the standard foundation for deep learning based WF methods.

Subsequent works extend the DF backbone in different directions. Var-CNN~\cite{bhat2018var} supplemented direction sequences with inter-packet timing and metadata, using a ResNet-based architecture to improve data efficiency and open-world performance. Tik-Tok~\cite{rahman2019tik} adopted the same CNN structure as DF but replaced the direction-only input with direction-multiplied-by-timestamp values, demonstrating that timing signals can be effectively exploited alongside directional features. Triplet Fingerprinting~\cite{sirinam2019triplet} introduced N-shot learning to improve portability and reduce data collection burden. Cherubin et al.~\cite{cherubin2022online} conducted the first evaluation using genuine Tor traffic as ground truth, providing a realistic open-world benchmark. NetAugment and NetCLR~\cite{bahramali2023realistic} proposed trace augmentation and self-supervised contrastive learning over direction sequences to enhance generalization across unobserved network conditions. A distinct representation choice appears in RF~\cite{shen2023subverting}, which constructs a Traffic Aggregation Matrix (TAM) by counting packet directions within fixed time slots, forming a 2D structure that tolerates timing perturbations introduced by defenses. LASERBEAK~\cite{mathews2024laserbeak} further introduced a Transformer-based approach with multi-channel feature representations including direction, timing, and size, demonstrating enhanced effectiveness under FRONT defenses.

Despite the diversity of these approaches in representation and architecture, all operate under the single-tab assumption, where each trace largely preserves the holistic fingerprint of one target website. This assumption is fundamentally challenged in multi-tab browsing, where traffic from multiple websites is interleaved, producing a qualitative change in problem structure that we characterize in this work as implicit traffic demixing.

\subsection{Multi-Tab and Multi-Label Website Fingerprinting}

\subsubsection*{Early Multi-Tab Work}

Multi-tab browsing fundamentally changes the WF problem structure. A single observed trace may contain overlapping flows from multiple websites, and the attacker must identify a set of visited sites rather than a single label. Xu et al.~\cite{xu2018multi} explicitly studied this setting and demonstrated that single-tab WF performance degrades severely when the single-page assumption is violated, motivating dedicated multi-tab attacks.

\subsubsection*{Sequence-Based Multi-Tab Methods}

Several multi-tab methods process traffic without explicit multi-level burst aggregation, instead relying on packet direction sequences or windowed encodings as their primary input representation. BAPM~\cite{guan2021bapm} generates a tab-aware representation from packet direction sequences and performs block division to separate concurrent page tabs as clearly as possible, using attention-based profiling to group blocks belonging to the same tab. TMWF~\cite{jin2023transformer} adopts DF's single-scale CNN as a local feature extractor and replaces DF's classification head with a Transformer encoder, demonstrating the utility of attention mechanisms for associating features across mixed-session traces. ARES'23~\cite{deng2023robust} advanced this line by explicitly formulating multi-tab WF as multi-label classification under a Transformer backbone with windowed packet-direction inputs, and contributed the closed-world and open-world benchmark datasets that have since become standard in the field. Despite differences in architecture, these methods share the absence of burst-level aggregation, and all rely on single-scale local feature encoding inherited from the DF architecture.

\subsubsection*{Aggregation-Based Multi-Tab Methods}

ARES'25~\cite{deng2025towards} departed from the above line by introducing Multi-Level Traffic Aggregation as its key contribution. The traffic trace is divided into fixed-size non-overlapping windows, and within each window both packet-level features (the sequence of packet directions) and burst-level features (burst count, average burst size, inter-burst intervals) are extracted. These per-window multi-level aggregated features are fed into an improved Transformer-based classifier. The structured local representation provided by Multi-Level Traffic Aggregation substantially improves robustness over sequence-based methods, and ARES'25 represents the current state of the art in multi-tab WF.

\subsubsection*{Limitations of Existing Methods}
Examining the above methods collectively reveals three structural limitations that have not been simultaneously addressed. First, all prior multi-tab methods rely on single-scale local feature extraction. Small convolution kernels capture fine-grained burst patterns but are sensitive to noise from concurrent flows, while large kernels model coarser structures but oversmooth the burst-level cues critical for fingerprint discrimination. Neither operating point adequately handles the multi-scale temporal diversity of multi-tab traffic, leaving (R1) unaddressed. Second, ARES'25's fixed non-overlapping segmentation fragments burst boundaries across adjacent windows, systematically discarding the inter-burst transition cues that are most informative for distinguishing concurrent sources, leaving (R2) unaddressed in existing fixed-window methods. Third, Transformer-based multi-tab methods~\cite{jin2023transformer,deng2023robust,deng2025towards} adopt absolute or learned positional encodings. However, the absolute positions of fragments from the same website are determined by unpredictable interleaving with concurrent flows, making absolute indices unreliable for fragment association. Relative temporal context is structurally more appropriate for the demixing problem, yet remains unexploited in existing approaches, leaving (R3) unaddressed. Collectively, these limitations indicate that multi-tab traffic demixing requires an architecture jointly designed to satisfy all three requirements, a goal that motivates the design of DEMUX.

\section{Threat Model}
In our threat model, clients use Tor or similar anonymous communication systems to conceal their online activity and may open $k$ browser tabs concurrently or at short intervals within a single browsing session, where $k$ is dynamic and unknown to the attacker. The resulting encrypted traffic originates from multiple target websites and overlaps temporally, preventing any individual website from presenting a complete, isolated traffic pattern. We further consider the deployment of fingerprint defense mechanisms such as WTF-PAD~\cite{juarez2015wtf}, FRONT~\cite{gong2020zero}, and TrafficSliver~\cite{de2020trafficsliver}, which operate at the client browser or Tor relay level to perturb traffic characteristics through techniques including dummy packet injection, adaptive padding, and multi-path traffic splitting.

As illustrated in Fig.~\ref{fig:tm}, we assume that the attacker is a passive eavesdropper positioned between the client and the Tor entry node, capable of capturing packet-level metadata for all outgoing and incoming traffic. Specifically, the attacker observes a sequence of packets, each represented as a direction-timestamp pair $(d_i, t_i)$, where $d_i \in \{+1, -1\}$ denotes the packet direction (outgoing or incoming) and $t_i$ denotes the arrival timestamp. No payload content is accessible. The attacker's objective is to infer the complete set of websites visited by the client within a session from this mixed, encrypted observation, thereby undermining the anonymity guarantees provided by the underlying network.

Our threat model builds upon the multi-tab WF setting established by ARES'23~\cite{deng2023robust}, whose benchmark datasets we directly adopt for no-defense evaluation. Compared to earlier formulations~\cite{guan2021bapm,jin2023transformer} that assume a fixed and known $k$, a single browser version, and traces collected exclusively from website homepages, ARES'23 relaxed these constraints by supporting dynamic tab counts ranging from 2 to 5, heterogeneous Tor Browser versions (10.x--13.x), and a more realistic crawling strategy. We adopt this setting without modification. The attacker must robustly recover the full set of visited websites from mixed traffic under unknown tab counts and heterogeneous browser environments, without any prior knowledge of $k$ or version-specific traffic characteristics.

Regarding defense scenarios, since existing defense mechanisms are designed for single-tab traffic, we follow the synthesis procedure of~\cite{jin2023transformer} to construct two-tab defense datasets by combining independently defended single-tab traces. Defense evaluation is therefore conducted exclusively in the 2-tab setting. For the no-defense setting, we evaluate across the full range of 2 to 5 concurrent tabs. Consistent with prior work~\cite{deng2023robust,jin2023transformer,deng2025towards}, we retain two standard evaluation scenarios. In the \emph{closed-world} scenario, the client visits only websites from the Alexa Top-100 monitored set and the attacker has full training coverage of all monitored sites. In the \emph{open-world} scenario, the client may visit arbitrary websites and the attacker possesses training samples drawn solely from the monitored subset.

\begin{figure*}[t] 
    \centering \includegraphics[width=\textwidth]{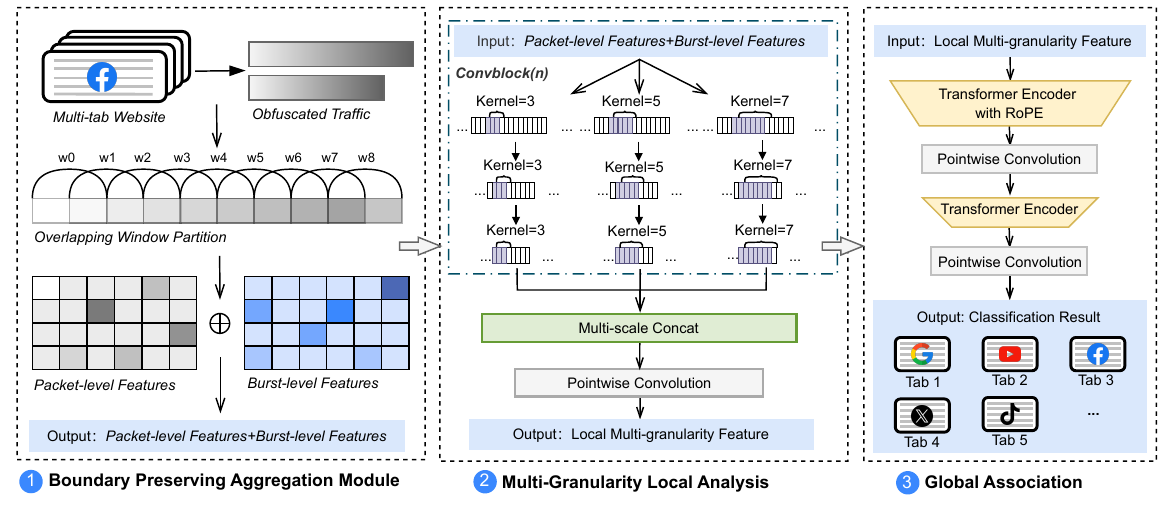}
    \caption{The DEMUX framework diagram includes the Boundary Preserving Aggregation Module, Multi-Granularity Local Analysis, and Global Association modules. It also provides detailed visualizations of the convolutional blocks (with kernel size = n) and the Transformer encoder, which incorporates a Multi-Headed Self-Attention mechanism enhanced with Rotary Positional Embedding (RoPE).}
    \label{fig:s}
\end{figure*}

\section{Framework}
\label{sec:framework}

This section presents the DEMUX framework. We formulate the multi-tab WF problem and provide an architectural overview (Section~\ref{sec:overview}), then detail the three components in Sections~\ref{sec:bm}--\ref{sec:ga}.

\subsection{Problem Formulation and Architecture Overview}
\label{sec:overview}

\noindent\textbf{Problem formulation.}
Let $\mathcal{T} = \{(d_i, t_i)\}_{i=1}^{N}$ denote an observed encrypted traffic trace of $N$ packets, where $d_i \in \{+1,-1\}$ encodes the packet direction (outgoing/incoming) and $t_i \in \mathbb{R}_{+}$ is the arrival timestamp. In a multi-tab browsing session, $\mathcal{T}$ is a temporally superimposed mixture of flows from $K$ concurrently visited websites, where $K$ is unknown to the attacker. The objective is to infer a binary label vector $\mathbf{y} = [y_1, \ldots, y_M]^\top \in \{0,1\}^M$ over $M$ monitored websites, where $y_m = 1$ if and only if website~$m$ was visited. This constitutes a \emph{multi-label classification} problem whose difficulty stems from two entangled factors: the input~$\mathcal{T}$ is an unstructured superposition of multiple sources, and both the number of sources~$K$ and their temporal boundaries are unknown.

\smallskip\noindent\textbf{Architecture overview.}
As motivated by the three structural requirements identified in Section~\ref{sec:introduction}, DEMUX comprises three sequentially composed modules, each targeting one requirement:

\begin{enumerate}
\item \textbf{Boundary Preserving Aggregation Module (Section~\ref{sec:bm})} partitions $\mathcal{T}$ into overlapping temporal windows and aggregates each window into a joint packet-level and burst-level feature vector, producing a sequence $\mathbf{X} \in \mathbb{R}^{L \times C}$ that preserves burst-boundary transition signals~(\textbf{R1}).

\item \textbf{Multi-Granularity Local Analysis (Section~\ref{sec:mgla})} processes $\mathbf{X}$ through a Multi-Scale Parallel CNN (MSP-CNN) with three parallel branches of heterogeneous kernel sizes, yielding a unified local feature map $\mathbf{H} \in \mathbb{R}^{L' \times d}$ that jointly encodes fine-grained burst patterns and coarse-grained periodic structures~(\textbf{R2}).

\item \textbf{Global Association (Section~\ref{sec:ga})} models long-range temporal dependencies via a two-stage Transformer encoder~\cite{vaswani2017attention} with Rotary Positional Embedding (RoPE)~\cite{su2024roformer}, associating dispersed fingerprint fragments from the same website across the full sequence~(\textbf{R3}).
\end{enumerate}

The complete architecture is illustrated in Fig.~\ref{fig:s}, and all hyperparameters are listed in Table~\ref{tab:model_parameters}.

\subsection{Boundary Preserving Aggregation Module}
\label{sec:bm}

\noindent\textbf{Motivation.}
We first define two key concepts. A \emph{burst} is a maximal consecutive sub-sequence of packets sharing the same direction. A \emph{burst boundary} is the transition point between two successive bursts with opposing directions. In multi-tab traffic, burst boundaries carry the most discriminative cross-source switching signals, as they mark the moments at which the dominant contributing website changes. Conventional fixed-size non-overlapping windows (length $W$, stride $W$), widely adopted in prior WF pipelines~\cite{ deng2025towards}, systematically fragment these boundaries: a burst straddling a window edge is split into two contextually isolated halves, destroying the switching evidence critical to \textbf{R1}.

\smallskip\noindent\textbf{Overlapping window partitioning.}
To address this limitation, the Boundary Preserving Aggregation Module replaces non-overlapping slicing with a sliding window of length~$W$ and stride~$\Delta < W$. Given a trace of total duration~$T$, the $k$-th window covers the time interval $[k\Delta,\; k\Delta + W)$. The total number of windows is
\begin{equation}
  L = \left\lfloor \frac{T - W}{\Delta} \right\rfloor + 1.
  \label{eq:num_windows}
\end{equation}
Since $\Delta < W$, every point in the trace is covered by $r = \lceil W/\Delta \rceil$ consecutive windows, guaranteeing that every burst boundary is fully contained within at least one window with sufficient context on both sides. In our implementation, $W = 20$\,ms and $\Delta = 10$\,ms, yielding a $50\%$ overlap ratio and $r = 2$.

\smallskip\noindent\textbf{Multi-level feature aggregation.}
Following the multi-level feature design introduced by ARES'25~\cite{ deng2025towards}, each window $w_k$ ($k = 1, \ldots, L$) extracts features at two complementary granularities:
\begin{itemize}
  \item \emph{Packet-level features}
    $\mathbf{p}_k \in \mathbb{R}^{C_p}$: the ordered sequence of
    packet directions $d_i \in \{+1,-1\}$ within $w_k$, capturing
    fine-grained directional dynamics.

  \item \emph{Burst-level features}
    $\mathbf{b}_k \in \mathbb{R}^{C_b}$: computed by grouping
    consecutive same-direction packets into bursts and extracting
    four structural descriptors---burst count, mean burst size,
    burst size variance, and mean inter-burst interval---that
    characterize the temporal rhythm of the window.
\end{itemize}
The two representations are concatenated into a unified window feature vector:
\begin{equation}
  \mathbf{x}_k =
    \bigl[\,\mathbf{p}_k \;\|\; \mathbf{b}_k\,\bigr]
    \in \mathbb{R}^{C}, \quad C = C_p + C_b,
  \label{eq:window_feat}
\end{equation}
and the module output is the sequence:
\begin{equation}
  \mathbf{X} =
    \bigl[\mathbf{x}_1,\; \mathbf{x}_2,\; \ldots,\;
    \mathbf{x}_L\bigr]^\top
    \in \mathbb{R}^{L \times C}.
  \label{eq:bpa_output}
\end{equation}
Packet-level features supply the directional evidence needed to discriminate burst shapes, while burst-level features supply the structural context needed to associate consecutive bursts from the same source. Together with overlapping partitioning, $\mathbf{X}$ constitutes a boundary-aware, structurally informative representation that directly satisfies~\textbf{R1}.

\smallskip\noindent\textbf{Plug-and-play applicability.}
The Boundary Preserving Aggregation Module is designed as a model-agnostic preprocessing component: it can replace the aggregation stage of any WF pipeline that relies on temporal or burst-derived features without modifying the downstream model. Its consistent improvements across diverse baselines are demonstrated in Section~\ref{sec:ablation}.

\subsection{Multi-Granularity Local Analysis}
\label{sec:mgla}

\noindent\textbf{Motivation.}
Given $\mathbf{X}$, the local analysis module must extract discriminative per-window and inter-window patterns. As discussed in Section~\ref{sec:introduction}, fragments from different websites coexist at heterogeneous temporal scales within a mixed trace: fine-grained burst patterns demand short receptive fields, while coarse-grained periodic loading rhythms require long receptive fields. Conventional single-kernel CNN backbones~\cite{sirinam2018deep, jin2023transformer} enforce a hard trade-off between these two regimes, making them structurally insufficient for~\textbf{R2}.

\smallskip\noindent\textbf{Multi-Scale Parallel CNN (MSP-CNN).}
MSP-CNN resolves this trade-off by deploying $B = 3$ independent convolutional branches with distinct kernel sizes $k_i \in \{3, 5, 7\}$:
\begin{equation}
  \mathbf{H}_i = \mathcal{R}_{k_i}(\mathbf{X})
    \in \mathbb{R}^{L' \times d_c}, \quad i \in \{1, 2, 3\},
  \label{eq:branches}
\end{equation}
where $\mathcal{R}_{k_i}(\cdot)$ denotes a branch consisting of stacked Residual 1D Convolutional Blocks (RCBs) with kernel size $k_i$, and $L'$ reflects temporal compression from pooling layers. Branch~1 ($k=3$) focuses on short-range burst-level patterns; Branch~2 ($k=5$) captures intermediate-range structures; Branch~3 ($k=7$) models coarse-grained periodic behaviors with reduced noise sensitivity. All kernel sizes are odd to enable symmetric zero-padding and center-aligned feature extraction.

Each RCB follows a standard residual design~\cite{he2016deep}:
\begin{equation}
  \mathrm{RCB}_{k}(\mathbf{z}) =
    \mathrm{BN}\!\bigl(\sigma\!\bigl(\mathrm{Conv}_{k}(\mathbf{z})
    \bigr)\bigr) + \mathbf{z},
  \label{eq:rcb}
\end{equation}
where $\mathrm{Conv}_{k}$ is a 1D convolution with kernel size~$k$, $\sigma(\cdot)$ is the activation function, and $\mathrm{BN}$ denotes batch normalization~\cite{ioffe2015batch}. The residual shortcut enables each branch to selectively amplify scale-specific patterns without redundantly relearning shared low-level statistics.

\smallskip\noindent\textbf{Multi-scale feature fusion.}
The three branch outputs are concatenated along the channel dimension:
\begin{equation}
  \mathbf{H}^{\mathrm{cat}} =
    \bigl[\,\mathbf{H}_1 \;\|\; \mathbf{H}_2 \;\|\;
    \mathbf{H}_3\,\bigr]
    \in \mathbb{R}^{L' \times 3d_c}.
  \label{eq:concat}
\end{equation}
Directly passing $\mathbf{H}^{\mathrm{cat}}$ to the subsequent Transformer would triple the channel dimensionality, increasing computation and introducing cross-branch redundancy. We therefore apply a Pointwise Convolution (PWConv), i.e., a $1 \times 1$ convolution operating exclusively along the channel dimension, to fuse and compress the multi-scale representations:
\begin{equation}
  \mathbf{H} = \mathrm{PWConv}(\mathbf{H}^{\mathrm{cat}})
    \in \mathbb{R}^{L' \times d},
  \label{eq:pwconv_fusion}
\end{equation}
where $3d_c = 768$ and $d = 256$ in our implementation (Table~\ref{tab:model_parameters}). PWConv performs cross-scale feature interaction and dimensionality reduction without altering the temporal structure, serving as a lightweight bottleneck that distills complementary scale-specific signals into a compact representation for global modeling.

\subsection{Global Association}
\label{sec:ga}

\noindent\textbf{Motivation.}
In multi-tab traffic, fingerprint cues from the same website are temporally dispersed throughout $\mathbf{H}$, interleaved with fragments from concurrent sources. Local CNN features alone cannot associate windows sharing a common source---this requires long-range temporal reasoning, motivating a Transformer-based global association module. We adopt a two-stage encoder design, where Stage~1 establishes relative positional alignment among windows (\textbf{R3}) and Stage~2 refines the globally associated representation in an expanded feature space before classification. An inter-stage pointwise projection decouples the two stages, keeping each architecturally focused on its respective purpose.

\smallskip\noindent\textbf{Stage~1: RoPE-enhanced Transformer encoder.}
Let $\mathbf{Z}^{(0)} = \mathbf{H} \in \mathbb{R}^{L' \times d}$. At each layer $\ell \in \{1,\ldots,L_1\}$, query, key, and value matrices are computed via learned linear projections:

\begin{align}
  \mathbf{Q}^{(\ell)} &= \mathbf{Z}^{(\ell-1)}\mathbf{W}_Q^{(\ell)}, \label{eq:query} \\
  \mathbf{K}^{(\ell)} &= \mathbf{Z}^{(\ell-1)}\mathbf{W}_K^{(\ell)}, \label{eq:key} \\
  \mathbf{V}^{(\ell)} &= \mathbf{Z}^{(\ell-1)}\mathbf{W}_V^{(\ell)}, \label{eq:value}
\end{align}
with $\mathbf{W}_Q^{(\ell)}, \mathbf{W}_K^{(\ell)}, \mathbf{W}_V^{(\ell)} \in \mathbb{R}^{d \times d}$.

Standard absolute positional encodings---whether sinusoidal~\cite{vaswani2017attention} or learned---assign positions by absolute index. In multi-tab traffic, this is unreliable because the absolute position of any website's fragment is determined by unpredictable interleaving with concurrent flows and varies across traces. We instead adopt Rotary Positional Embedding (RoPE)~\cite{su2024roformer}, which encodes positions by rotating query and key vectors in pairs of embedding subspaces. For position index~$m$, the $j$-th two-dimensional subspace is rotated by angle $m\theta_j$ with $\theta_j = 10000^{-2j/d}$:
\begin{equation}
  \tilde{\mathbf{Q}}^{(\ell)} =
    \mathrm{RoPE}\!\bigl(\mathbf{Q}^{(\ell)}\bigr), \qquad
  \tilde{\mathbf{K}}^{(\ell)} =
    \mathrm{RoPE}\!\bigl(\mathbf{K}^{(\ell)}\bigr).
  \label{eq:rope}
\end{equation}
The key property is that the resulting attention score depends only on relative positional offsets:
\begin{equation}
  \bigl\langle
    \tilde{\mathbf{Q}}^{(\ell)}_m,\;
    \tilde{\mathbf{K}}^{(\ell)}_n
  \bigr\rangle
  = f(m - n),
  \label{eq:rope_relative}
\end{equation}
where $f$ is a function of the relative displacement $(m-n)$ alone. This property directly addresses~\textbf{R3}: the attention score between two windows reflects their relative temporal distance, enabling the model to associate fragments from the same website regardless of their absolute positions in the mixed trace.

Multi-head self-attention with $n_h$ heads and per-head dimension $d_h = d/n_h$ is computed as
\begin{equation}
  \mathbf{Z}^{(\ell)} =
    \mathrm{Concat}_{h=1}^{n_h}\!\left(
      \mathrm{Softmax}\!\left(
        \frac{\tilde{\mathbf{Q}}^{(\ell)}_h\,
              \tilde{\mathbf{K}}^{(\ell)\top}_h}
             {\sqrt{d_h}}
      \right)\mathbf{V}^{(\ell)}_h
    \right)\mathbf{W}_O^{(\ell)},
  \label{eq:mhsa}
\end{equation}
where $\mathbf{W}_O^{(\ell)}\in\mathbb{R}^{d\times d}$. Each attention sub-layer is followed by a position-wise feed-forward network (FFN) with hidden dimension $4d$, and both sub-layers employ residual connections and layer normalization~\cite{ba2016layer}. After $L_1$ layers, Stage~1 outputs $\mathbf{Z}^{(1)} \in \mathbb{R}^{L' \times d}$.

\smallskip\noindent\textbf{Inter-stage feature expansion.}
A learnable pointwise projection expands each token from dimension $d$ to $d' > d$ before Stage~2:
\begin{equation}
  \mathbf{Z}^{(2)} = \mathrm{PWConv}\bigl(\mathbf{Z}^{(1)}\bigr)
    \in \mathbb{R}^{L' \times d'},
  \label{eq:interstage}
\end{equation}
where $d = 256$ and $d' = 384$. This expansion increases per-token representational capacity for Stage~2 refinement without perturbing the relative positional structure encoded by RoPE in Stage~1.

\smallskip\noindent\textbf{Stage~2: Refinement encoder.}
Stage~2 applies $L_2$ standard Transformer encoder layers (without RoPE) on $\mathbf{Z}^{(2)}$ with hidden dimension $d'$:
\begin{equation}
  \mathbf{Z}^{(3)} =
    \mathrm{TransformerEncoder}_{L_2,\,d'}\!\bigl(
      \mathbf{Z}^{(2)}
    \bigr)
    \in \mathbb{R}^{L' \times d'}.
  \label{eq:stage2}
\end{equation}
Each layer follows the same residual self-attention and FFN structure as Stage~1, with FFN dimension $4d'$.

\smallskip\noindent\textbf{Classification head.}
The sequence $\mathbf{Z}^{(3)}$ must be aggregated into a fixed-size vector for classification. Flattening all $L'$ position vectors inflates parameter count and propagates positional noise; plain average pooling suppresses discriminative variation across windows. We adopt an \emph{up-projection then pooling} strategy: a linear projection $\mathbf{W}_{\mathrm{up}} \in \mathbb{R}^{d' \times d''}$ expands each position-wise feature before averaging:
\begin{equation}
  \mathbf{z} =
    \frac{1}{L'}\sum_{l=1}^{L'}
    \mathbf{Z}^{(3)}_l\,\mathbf{W}_{\mathrm{up}}
    \in \mathbb{R}^{d''},
  \label{eq:pooling}
\end{equation}
where $d'' = 1024$. Increasing per-token capacity before pooling improves robustness to the alignment variability inherent in overlapping segmentation of multi-tab traffic (confirmed in Section~\ref{subsec:sensitivity}).

The pooled vector $\mathbf{z}$ is passed to an MLP classifier with sigmoid activation:
\begin{equation}
  \hat{\mathbf{y}} = \sigma\!\bigl(\mathrm{MLP}(\mathbf{z})\bigr)
    \in [0,1]^M,
  \label{eq:output}
\end{equation}
where $\hat{y}_m$ is the predicted probability that website~$m$ was visited. The model is trained end-to-end by minimizing the binary cross-entropy loss:
\begin{equation}
  \mathcal{L} = \mathrm{BCE}(\hat{\mathbf{y}},\; \mathbf{y}).
  \label{eq:bce}
\end{equation}

\smallskip\noindent\textbf{Summary.}
The three modules of DEMUX form a tightly coupled pipeline derived directly from the structural requirements of multi-tab traffic demixing: the Boundary Preserving Aggregation Module ensures burst-boundary integrity~(\textbf{R1}), MSP-CNN provides multi-scale local perception~(\textbf{R2}), and the RoPE-enhanced Transformer enables relative temporal association of dispersed fragments~(\textbf{R3}). Their interactions are systematically analyzed in Sections~\ref{sec:ablation} and~\ref{subsec:sensitivity}.

\section{Experiments}
\label{sec:experiments}

In this section, we conduct an extensive experimental evaluation of \textsc{DEMUX} to assess its effectiveness and robustness in realistic multi-tab website fingerprinting
scenarios.
Our evaluation compares \textsc{DEMUX} against state-of-the-art (SOTA) baselines
across a broad range of settings, including closed-world and open-world
scenarios, dynamic tab configurations, and traffic traces protected by
representative website fingerprinting defense mechanisms.
Beyond overall detection performance, we further conduct ablation analysis and
sensitivity analysis to systematically analyze the contribution of key model
components and architectural design choices.

\subsection{Experimental Setup}

\subsubsection{Datasets}

We evaluate DEMUX on datasets spanning four complementary scenarios.

\textbf{Multi-tab datasets.}
We adopt the closed-world and open-world benchmark datasets from Deng~\emph{et al.}~\cite{deng2023robust}, collected over Tor without any traffic defense.
In the \emph{closed-world} setting, the Alexa top-100 websites serve as monitored classes; each trace interleaves 2--5 concurrently loaded tabs drawn uniformly from this set, yielding 58{,}000 traces per tab configuration.
In the \emph{open-world} setting, each $N$-tab trace combines $N{-}1$ monitored websites with one unmonitored site drawn from the Alexa top-20{,}000, yielding 64{,}000 traces per configuration.

\textbf{Defense datasets.}
To evaluate robustness under traffic obfuscation, we construct 2-tab 
datasets protected by three representative defenses: 
WTF-PAD~\cite{juarez2015wtf}, Front~\cite{gong2020zero}, and 
TrafficSliver~\cite{de2020trafficsliver}. Since these defenses target 
single-tab traffic, we follow the synthesis procedure of 
Jin~\emph{et al.}~\cite{jin2023transformer} and compose multi-tab traces 
by overlaying independently defended single-tab captures using the 
authors' official implementations. WTF-PAD injects dummy packets during 
idle periods to obscure inter-burst timing gaps; Front concentrates dummy 
packet injection at the beginning of each trace using a Rayleigh 
distribution to obfuscate the feature-rich trace front with zero latency 
overhead; TrafficSliver splits traffic across multiple Tor entry nodes so 
that an adversary at any single node can observe only a partial fraction 
of packets, without introducing artificial delays or dummy traffic.

\textbf{Dynamic-tab dataset.}
To assess generalization under unknown tab counts, we construct a mixed training set by sampling 15{,}000 traces from each closed-world 2--5-tab split (60{,}000 traces total). Models trained on this set are evaluated separately on each fixed-tab test split.

\subsubsection{Baselines}

We compare DEMUX against nine representative methods spanning two architectural families.

\textbf{CNN-based:} DF~\cite{sirinam2018deep}, Var-CNN~\cite{bhat2018var}, Tik-Tok~\cite{rahman2019tik}, RF~\cite{shen2023subverting}, and NetCLR~\cite{bahramali2023realistic}.

\textbf{Transformer-based:} BAPM~\cite{guan2021bapm}, TMWF~\cite{jin2023transformer}, ARES'23~\cite{deng2023robust}, and ARES'25~\cite{deng2025towards}.

BAPM, TMWF, ARES'23, and ARES'25 are natively designed for multi-tab classification and trained with sigmoid outputs under binary cross-entropy (BCE) loss. The CNN-based single-tab methods are adapted by replacing their softmax heads with sigmoid layers and retraining under the same BCE objective, enabling each to produce independent site-level probabilities. All baselines are re-implemented and re-trained within a unified preprocessing and training pipeline to ensure fair comparison.

\subsubsection{Evaluation Metrics}
\label{sec:metrics}

Following standard practice in multi-tab WF evaluation~\cite{deng2023robust,deng2025towards}, we report three complementary metrics—AUC~\cite{ling2003auc}, P@$K$, and MAP@$K$—each computed per instance and averaged over the test set. Let $\mathbf{y} \in \{0,1\}^{C}$ be the ground-truth label vector and $\hat{\mathbf{y}}$ the corresponding predicted scores.

\textbf{AUC} is computed per site via one-versus-all ROC curves and averaged, providing a threshold-free measure robust to the severe class imbalance inherent in multi-tab traffic.

\textbf{P@$K$} measures the fraction of ground-truth sites recovered within the top-$K$ predictions:
\begin{equation}\label{eq:p-at-k}
\mathrm{P@}K = \frac{1}{K}\sum_{l\in\mathrm{Top}_K(\hat{\mathbf{y}})} y_l.
\end{equation}

\textbf{MAP@$K$} further rewards higher-ranked correct labels by accumulating prefix precisions:
\begin{equation}\label{eq:map-at-k}
\mathrm{MAP@}K = \frac{1}{K}\sum_{k=1}^{K} \mathrm{P@}k.
\end{equation}

AUC captures global discrimination, while P@$K$ and MAP@$K$ directly reflect the attacker's practical success in deanonymizing the most likely visited sites.

\subsubsection{Implementation}
\label{subsubsec:implementation}

DEMUX is implemented in PyTorch. Each dataset is partitioned once into 
train/validation/test subsets at an 8:1:1 ratio with a fixed random seed 
of 2025, governing data splits, weight initialization, and all stochastic 
components. For baseline methods, we adopt the open-source implementation 
framework of Deng~\emph{et al.}~\cite{deng2024wflib}, retaining each 
model's original training configuration as provided in the library to preserve the 
performance reported in the respective original works. DEMUX itself is trained with the AdamW 
optimizer~\cite{loshchilov2018decoupled} with a weight decay of 
$5\times10^{-3}$, a linear warm-up over the first 10 epochs from 
$2\times10^{-4}$ to $2\times10^{-3}$ followed by cosine annealing, 
for a total of 260 epochs at a batch size of 512. All experiments are 
conducted on a single NVIDIA RTX~4090 GPU. Detailed architectural and 
training hyperparameters of DEMUX are listed in 
Table~\ref{tab:model_parameters}.

\begin{table}[t]
\centering
\caption{Architecture and Training Configuration of DEMUX.}
\label{tab:model_parameters}
\resizebox{\linewidth}{!}{%
\begin{tabular}{llc}
\toprule
\textbf{Module} & \textbf{Hyperparameter} & \textbf{Value} \\
\midrule
\multicolumn{3}{l}{\textit{Boundary Preserving Aggregation Module (BM)}} \\
& Window size / stride & 20\,ms / 10\,ms \\
& Feature channels ($C$) & 8 \\
\midrule
\multicolumn{3}{l}{\textit{Multi-Scale Parallel CNN (MSP-CNN)}} \\
& Kernel sizes & \{3, 5, 7\} \\
& Channel progression & $8{\to}32{\to}64{\to}128{\to}256$ \\
& Pooling (kernel/stride) & 8/4 $\times$ 4 stages \\
& Fusion (pointwise conv) & $768 \to 256$ \\
\midrule
\multicolumn{3}{l}{\textit{Global Association — Stage 1}} \\
& Layers / heads / dim & 2 / 8 / 256 \\
& FFN dim & 1024 \\
& Positional encoding & RoPE (base $10^4$) \\
\midrule
\multicolumn{3}{l}{\textit{Global Association — Stage 2 }} \\
& Inter-stage projection & $256 \to 384$ \\
& Layers / heads / dim & 2 / 8 / 384 \\
& FFN dim & 1536 \\
\midrule
\multicolumn{3}{l}{\textit{Classification Head}} \\
& Aggregation & Avg-pool + linear ($384{\to}1024$) \\
& Output & Sigmoid MLP ($N_\text{cls}$ classes) \\
\midrule
\multicolumn{3}{l}{\textit{Training}} \\
& Optimizer / weight decay & AdamW / $5\times10^{-3}$ \\
& LR schedule (max / warm-up) & Cosine / $2\times10^{-3}$ / 10 epochs \\
& Batch size / epochs & 512 / 260 \\
& Hardware & NVIDIA RTX 4090 \\
\bottomrule
\end{tabular}}
\end{table}


\vspace{4pt}
\begin{table*}[tb]
\centering
\caption{Closed‑world performance of multi‑tab WF attacks.}
\label{tab:closedworld}
\resizebox{\linewidth}{!}{%
\begin{tabular}{lcccccccccccc}
\toprule
\multirow{2}{*}{\textbf{Method}} 
& \multicolumn{3}{c}{\textbf{2‑tab}} & \multicolumn{3}{c}{\textbf{3‑tab}} 
& \multicolumn{3}{c}{\textbf{4‑tab}} & \multicolumn{3}{c}{\textbf{5‑tab}} \\
\cmidrule(lr){2-4} \cmidrule(lr){5-7} \cmidrule(lr){8-10} \cmidrule(lr){11-13}
& AUC & P@2 & MAP@2 & AUC & P@3 & MAP@3 & AUC & P@4 & MAP@4 & AUC & P@5 & MAP@5 \\
\midrule
BAPM        & 0.935 & 0.529 & 0.625 & 0.867 & 0.377 & 0.478 & 0.839 & 0.349 & 0.446 & 0.793 & 0.299 & 0.385 \\
NetCLR      & 0.943 & 0.590 & 0.688 & 0.872 & 0.425 & 0.543 & 0.846 & 0.387 & 0.498 & 0.797 & 0.325 & 0.426 \\
DF          & 0.944 & 0.601 & 0.712 & 0.861 & 0.421 & 0.560 & 0.831 & 0.373 & 0.512 & 0.772 & 0.300 & 0.423 \\
RF          & 0.950 & 0.643 & 0.752 & 0.876 & 0.489 & 0.649 & 0.840 & 0.427 & 0.604 & 0.779 & 0.338 & 0.499 \\
Tik-Tok     & 0.957 & 0.647 & 0.754 & 0.872 & 0.443 & 0.588 & 0.837 & 0.383 & 0.531 & 0.781 & 0.306 & 0.428 \\
Var-CNN     & 0.971 & 0.726 & 0.809 & 0.923 & 0.562 & 0.697 & 0.866 & 0.429 & 0.566 & 0.786 & 0.326 & 0.468 \\
TMWF        & 0.973 & 0.740 & 0.805 & 0.936 & 0.590 & 0.679 & 0.933 & 0.619 & 0.701 & 0.908 & 0.548 & 0.633 \\
ARES'23   & 0.987 & 0.832 & 0.880 & 0.979 & 0.778 & 0.846 & 0.974 & 0.774 & 0.841 & 0.966 & 0.732 & 0.805 \\
ARES'25   & 0.994 & 0.900 & 0.936 & 0.990 & 0.864 & 0.913 & 0.989 & 0.887 & 0.926 & 0.986 & 0.851 & 0.899 \\
\textbf{DEMUX} 
            & \textbf{0.997} & \textbf{0.926} & \textbf{0.953} 
            & \textbf{0.996} & \textbf{0.917} & \textbf{0.947} 
            & \textbf{0.995} & \textbf{0.931} & \textbf{0.953} 
            & \textbf{0.996} & \textbf{0.943} & \textbf{0.961} \\
\bottomrule
\end{tabular}}
\end{table*}


\vspace{4pt}
\begin{table*}[htbp]
\centering
\caption{Open‑world performance of multi‑tab WF attacks.}
\label{tab:openworld}
\resizebox{\linewidth}{!}{%
\begin{tabular}{lcccccccccccc}
\toprule
\multirow{2}{*}{\textbf{Method}} 
& \multicolumn{3}{c}{\textbf{2‑tab}} & \multicolumn{3}{c}{\textbf{3‑tab}} 
& \multicolumn{3}{c}{\textbf{4‑tab}} & \multicolumn{3}{c}{\textbf{5‑tab}} \\
\cmidrule(lr){2-4} \cmidrule(lr){5-7} \cmidrule(lr){8-10} \cmidrule(lr){11-13}
& AUC & P@2 & MAP@2 & AUC & P@3 & MAP@3 & AUC & P@4 & MAP@4 & AUC & P@5 & MAP@5 \\
\midrule
BAPM        & 0.932 & 0.515 & 0.610 & 0.868 & 0.379 & 0.483 & 0.837 & 0.346 & 0.443 & 0.799 & 0.308 & 0.397 \\
NetCLR      & 0.940 & 0.575 & 0.674 & 0.874 & 0.428 & 0.546 & 0.846 & 0.382 & 0.495 & 0.791 & 0.332 & 0.442 \\
DF          & 0.941 & 0.577 & 0.687 & 0.861 & 0.424 & 0.565 & 0.827 & 0.374 & 0.511 & 0.780 & 0.315 & 0.446 \\
RF          & 0.948 & 0.639 & 0.749 & 0.880 & 0.497 & 0.656 & 0.841 & 0.430 & 0.606 & 0.788 & 0.354 & 0.520 \\
Tik-Tok     & 0.955 & 0.631 & 0.736 & 0.874 & 0.451 & 0.600 & 0.835 & 0.379 & 0.521 & 0.786 & 0.317 & 0.449 \\
Var-CNN     & 0.969 & 0.705 & 0.791 & 0.924 & 0.567 & 0.703 & 0.829 & 0.398 & 0.549 & 0.776 & 0.322 & 0.460 \\
TMWF        & 0.969 & 0.697 & 0.761 & 0.944 & 0.616 & 0.706 & 0.929 & 0.601 & 0.683 & 0.905 & 0.542 & 0.631 \\
ARES'23   & 0.985 & 0.806 & 0.859 & 0.976 & 0.770 & 0.844 & 0.973 & 0.774 & 0.841 & 0.931 & 0.587 & 0.702 \\
ARES'25   & 0.992 & 0.879 & 0.920 & 0.990 & 0.868 & 0.917 & 0.988 & 0.875 & 0.918 & 0.988 & 0.869 & 0.911 \\
\textbf{DEMUX } 
            & \textbf{0.996} & \textbf{0.913} & \textbf{0.944} 
            & \textbf{0.995} & \textbf{0.917} & \textbf{0.948} 
            & \textbf{0.996} & \textbf{0.931} & \textbf{0.954} 
            & \textbf{0.998} & \textbf{0.951} & \textbf{0.966} \\
\bottomrule
\end{tabular}}
\end{table*}

\subsection{Closed-World Evaluation}
\label{subsec:closedworld}

Table~\ref{tab:closedworld} reports closed-world performance as the 
number of concurrent tabs increases from 2 to 5. \textsc{DEMUX} 
consistently outperforms all baselines across all metrics, and its 
margin over the strongest baseline, ARES'25, widens with tab count. 
In terms of P@$K$, ARES'25 degrades from 0.900 at 2-tab to 0.851 at 
5-tab, a drop of 4.9 percentage points, whereas \textsc{DEMUX} improves 
from 0.926 to 0.943 over the same range, expanding the absolute margin 
from 2.6 to over 9 points. MAP@$K$ follows the same pattern: ARES'25 
declines by more than 3.7 points cumulatively, while \textsc{DEMUX} 
remains stable with a degradation below 1 point. AUC tells a similar 
story—\textsc{DEMUX} holds at 0.997--0.996 across all settings, 
fluctuating by less than 0.001, whereas ARES'25 drops by approximately 
0.008. These results indicate that \textsc{DEMUX} not only achieves 
higher absolute performance but also degrades significantly more slowly 
as traffic mixing intensifies, confirming the effectiveness of its 
architectural design under increasingly challenging multi-tab conditions.

\subsection{Open-World Evaluation}
\label{subsec:openworld}

Table~\ref{tab:openworld} reports results in the open-world setting, 
where each trace contains one unmonitored site, substantially increasing 
noise and class imbalance. The performance trends observed in the 
closed-world setting are preserved and amplified here. ARES'25 exhibits 
a modest but consistent decline in P@$K$ from 0.879 to 0.869 as tabs 
increase from 2 to 5, whereas \textsc{DEMUX} improves from 0.913 to 
0.951, widening the absolute gap from 3.4 to over 8 points. MAP@$K$ 
follows the same direction: ARES'25 drops by nearly 1 point, while 
\textsc{DEMUX} rises steadily from 0.944 to 0.966. In the most 
challenging 5-tab setting, \textsc{DEMUX} attains an AUC of 0.998, 
surpassing ARES'25 by approximately 1 percentage point—a substantial 
margin at this performance level. Taken together, these results 
demonstrate that \textsc{DEMUX} maintains strong discriminability in the 
presence of unmonitored traffic and heavy flow interleaving, where 
existing methods show measurable degradation.

\begin{table*}[tb]
  \centering
  \caption{Results under three website-fingerprinting defenses (2-tab). The defense datasets were synthesized using the TMWF-provided synthesis code to convert single-label traces into two-label mixtures.}
  \label{tab:defense}
  \resizebox{\linewidth}{!}{%
  \begin{tabular}{l|ccc|ccc|ccc}
    \toprule
    \multirow{2}{*}{\textbf{Method}} &
    \multicolumn{3}{c|}{\textbf{WTF-PAD}} & \multicolumn{3}{c|}{\textbf{Front}} & \multicolumn{3}{c}{\textbf{TrafficSliver}} \\
    & AUC & P@2 & MAP@2 & AUC & P@2 & MAP@2 & AUC & P@2 & MAP@2 \\
    \midrule
    BAPM      & 0.941 & 0.562 & 0.671 & 0.897 & 0.414 & 0.506 & 0.795 & 0.286 & 0.363 \\
    NetCLR    & 0.945 & 0.672 & 0.784 & 0.914 & 0.554 & 0.671 & 0.849 & 0.396 & 0.499 \\
    DF        & 0.954 & 0.698 & 0.810 & 0.925 & 0.566 & 0.695 & 0.853 & 0.401 & 0.506 \\
    RF        & 0.959 & 0.769 & 0.858 & 0.955 & 0.743 & 0.838 & 0.956 & 0.702 & 0.791 \\
    Tik-Tok   & 0.966 & 0.751 & 0.850 & 0.945 & 0.629 & 0.759 & 0.938 & 0.570 & 0.682 \\
    Var-CNN   & 0.976 & 0.792 & 0.876 & 0.883 & 0.429 & 0.513 & 0.986 & 0.826 & 0.889 \\
    TMWF      & 0.992 & 0.869 & 0.919 & 0.981 & 0.792 & 0.866 & 0.864 & 0.399 & 0.497 \\
    ARES'23   & 0.986 & 0.881 & 0.930 & 0.972 & 0.808 & 0.881 & 0.845 & 0.429 & 0.533 \\
    ARES'25   & 0.997 & 0.951 & 0.973 & 0.997 & 0.949 & 0.971 & 0.995 & 0.915 & 0.949 \\
    \textbf{DEMUX } & \textbf{0.998} & \textbf{0.959} & \textbf{0.977} & \textbf{0.998} & \textbf{0.962} & \textbf{0.979} & \textbf{0.996} & \textbf{0.940} & \textbf{0.964} \\
    \bottomrule
  \end{tabular}
  }
\end{table*}

\subsection{Defense Robustness}
\label{subsec:defenses}

Table~\ref{tab:defense} reports performance under three representative 
defenses in the synthesized 2-tab setting. \textsc{DEMUX} achieves the 
highest AUC, P@2, and MAP@2 under all three defenses, and its margin 
over ARES'25 grows with the severity of the defense: 0.1, 0.8, and 0.4 
points under WTF-PAD; 0.1, 1.3, and 0.8 points under Front; and 0.1, 
2.5, and 1.5 points under TrafficSliver, respectively.

The more informative signal lies in how methods respond differently to 
each defense type. Under WTF-PAD, which targets inter-burst timing gaps, 
most methods maintain relatively strong performance, as directional 
patterns remain largely intact. Under Front, which obfuscates the 
feature-rich trace front, TMWF and ARES'23 degrade noticeably in P@2 
to 0.792 and 0.808, suggesting that their feature extraction is 
sensitive to front-loaded trace perturbation. TrafficSliver, which 
restricts each entry-node observer to only a partial fraction of 
packets, proves most disruptive: TMWF collapses to a P@2 of 0.399 
and ARES'23 to 0.429, while \textsc{DEMUX} retains a P@2 of 0.940, 
exceeding the next-best competitor by 2.5 points.

Two exceptions are worth noting. RF maintains competitive performance 
across all three defenses—particularly under TrafficSliver, where its 
P@2 of 0.702 substantially exceeds other CNN-based methods. This 
resilience is attributable to its TAM representation, which aggregates 
directional packet counts over fixed time slots and thus tolerates 
partial packet loss introduced by traffic splitting. Var-CNN similarly 
shows unexpectedly strong performance under TrafficSliver, achieving a 
P@2 of 0.826, despite its comparatively modest results in the 
no-defense setting.

\textsc{DEMUX} remains the top-ranked method across all nine 
metric--defense combinations. Its consistent advantage under 
TrafficSliver in particular, where incomplete packet observations 
break both local burst patterns and global temporal structure, 
suggests that its overlapping-window representation and multi-scale 
feature extraction provide complementary robustness that no single 
architectural choice in competing methods achieves alone.

\begin{figure*}[tb]
    \centering \includegraphics[width=\textwidth]{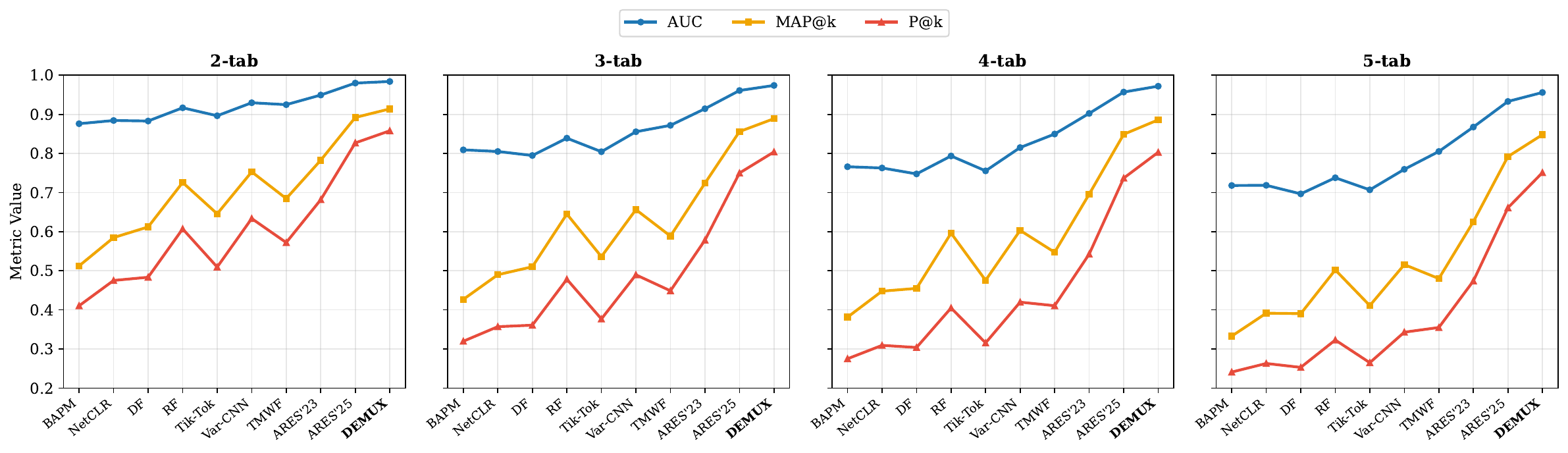}
    \caption{Dynamic-tab evaluation on the closed-world dataset. 
Models are trained on a mixed 2--5-tab training set and 
evaluated separately on each fixed-tab test split.}
    \label{fig:dynamic}
\end{figure*}

\subsection{Dynamic Tab Evaluation}
\label{subsec:dynamic}

In practice, the number of concurrently opened tabs is unknown and 
varies across browsing sessions. We evaluate whether \textsc{DEMUX} can 
handle this uncertainty by training on a mixed dataset of 60{,}000 
traces sampled uniformly from the closed-world 2--5-tab splits at 
15{,}000 traces per configuration, and evaluating separately on each 
fixed-tab test set. Figure~\ref{fig:dynamic} shows that \textsc{DEMUX} 
achieves the highest AUC, P@$K$, and MAP@$K$ across all twelve 
metric--setting pairs. Notably, its advantage over ARES'25 is maintained 
even in the 5-tab setting, where mixed training is most challenging due 
to higher label density and stronger inter-flow interference. These 
results indicate that \textsc{DEMUX} learns a unified representation that 
remains discriminative across varying tab compositions, rather than 
overfitting to the statistical properties of any single configuration.

\subsection{Cross-Configuration Generalization}
\label{subsec:generalization}

\begin{figure}[tb] 
    \centering \includegraphics[width=0.5\textwidth]{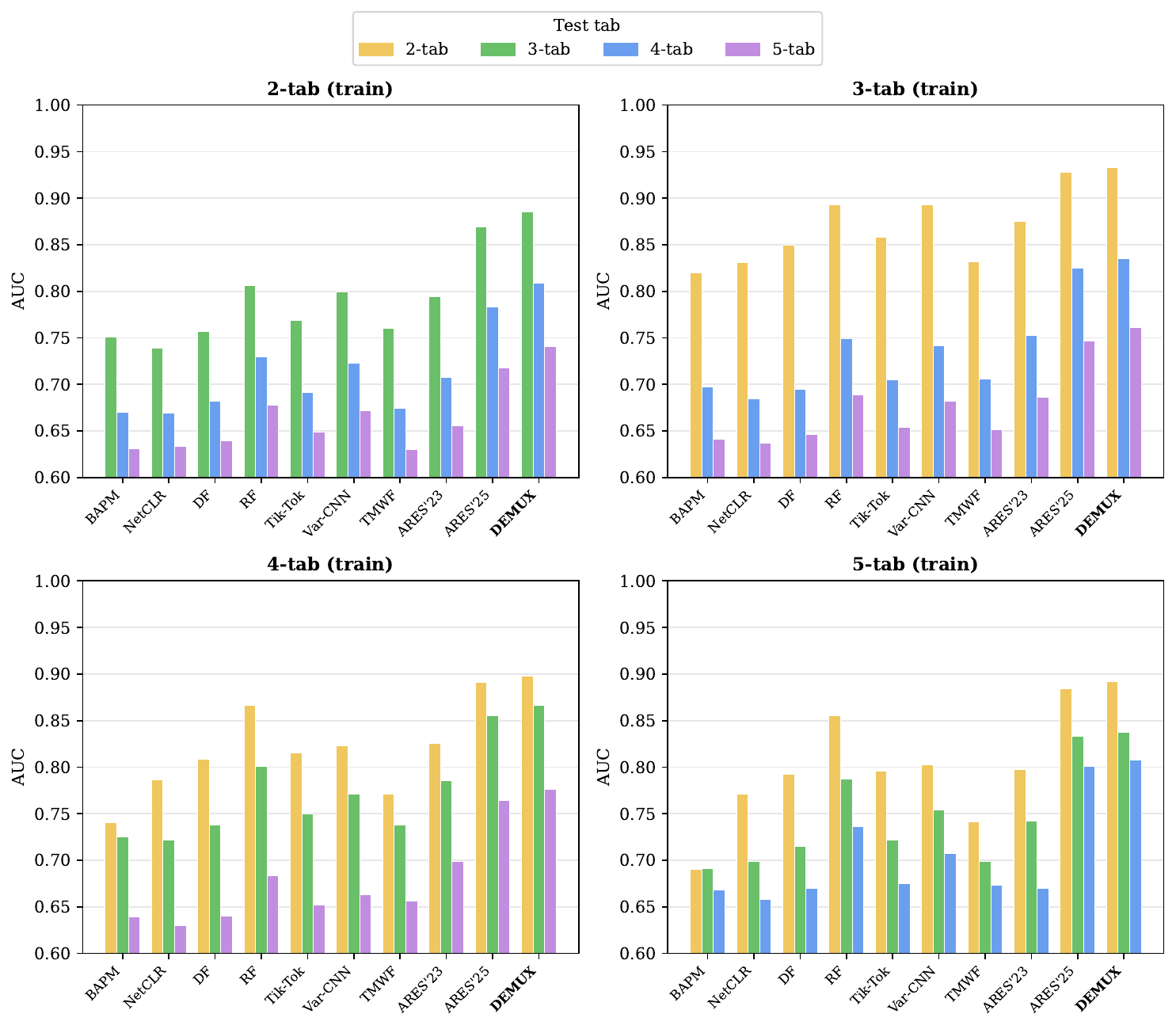}
    \caption{Cross-configuration generalization (AUC) on the 
closed-world dataset. Each subplot trains on one tab count 
and evaluates on all four.}
    \label{fig:generalization}
\end{figure}

The dynamic evaluation above trains on all tab counts simultaneously. 
Here we ask a strictly harder question: how well does a model trained 
exclusively on one tab configuration transfer to unseen tab counts? For 
each of the four closed-world configurations spanning 2 to 5 tabs, we 
train a dedicated model and evaluate it on the remaining three, yielding 
twelve train--test permutations in total. Since P@$K$ and MAP@$K$ 
require knowledge of the true label count, we report AUC only. As shown 
in Figure~\ref{fig:generalization}, \textsc{DEMUX} achieves the best AUC 
in all twelve permutations, with margins over the strongest baseline 
ranging from 0.4 to 2.6 percentage points. The largest gains appear in 
high-to-low transfer settings, where a model trained on 5-tab traces is 
evaluated on 2-tab or 3-tab data, and \textsc{DEMUX}'s richer multi-scale 
representations generalise more effectively to simpler traffic mixtures. 
Taken together with the dynamic evaluation, these results confirm that 
\textsc{DEMUX}'s robustness is not contingent on prior knowledge of tab 
count distribution, a realistic constraint in practical WF deployment 
scenarios.

\subsection{Effectiveness of the Boundary Preserving Aggregation Module}
\label{subsec:effectivenessOfOverlapping}

To quantify the contribution of Boundary Preserving Aggregation Module(BM) as a plug-and-play component, we 
integrate it into three representative baselines under the open-world 
5-tab setting. For DF and TMWF, the \textsc{+BM} variants replace 
their raw direction sequences with BM's multi-level representation. 
For ARES'25, \textsc{+BM} replaces its Multi-Level Traffic Aggregation with BM. We 
additionally include DEMUX-Dir, a direction-only ablation variant of 
\textsc{DEMUX} that uses raw packet directions without timestamp or 
burst-level information.

\begin{figure}[tb]
  \centering
  \includegraphics[width=\linewidth]{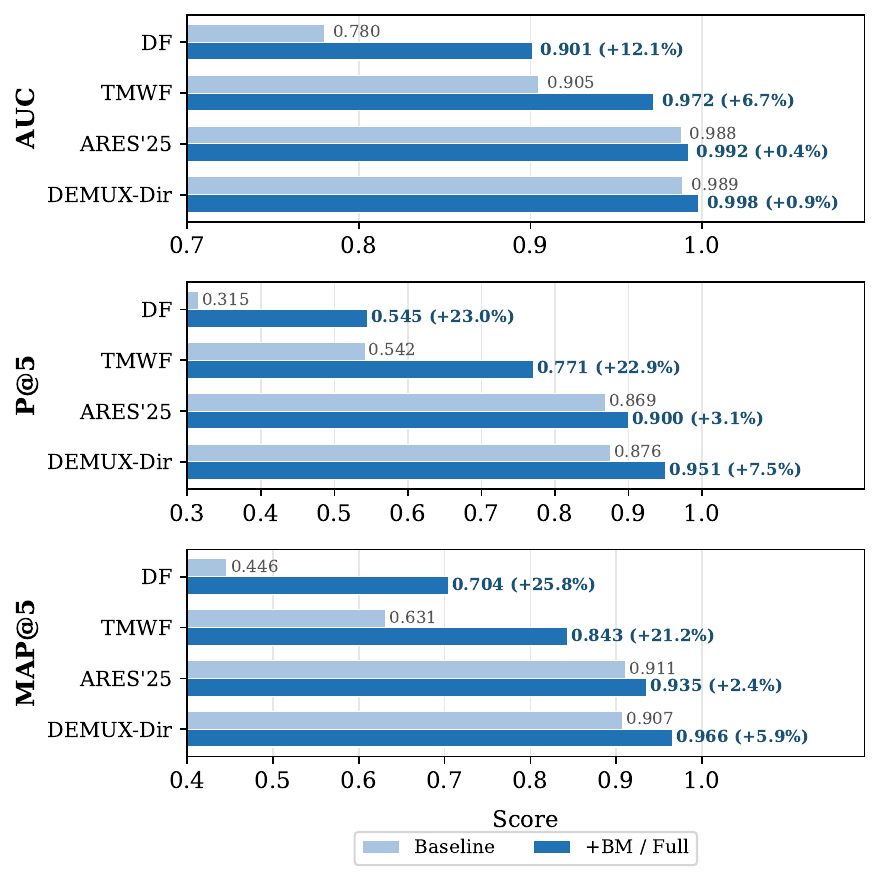}
  \caption{Effectiveness of the Boundary Preserving Aggregation Module 
  as a plug-and-play component in the open-world 5-tab setting. 
  Light bars denote baseline models; dark bars denote variants 
  with BM integrated (DEMUX-Dir vs.\ full DEMUX).}
  \label{fig:bm}
\end{figure}

Figure~\ref{fig:bm} shows that BM consistently improves all three 
baselines. The gains are largest for DF, where AUC rises from 0.780 
to 0.901 and P@5 from 0.315 to 0.545, reflecting the combined effect 
of boundary preservation and the richer multi-level input. TMWF 
improves similarly, with AUC from 0.905 to 0.972 and P@5 from 0.542 
to 0.771, confirming that cleaner local representations complement 
Transformer-based global modeling. Even ARES'25, which already 
employs structured multi-level aggregation, gains meaningfully from 
BM substitution, with P@5 rising from 0.869 to 0.900.

Comparing DEMUX-Dir and full \textsc{DEMUX} reveals that direction 
patterns alone yield a competitive AUC of 0.989, but incorporating 
BM's multi-level representation raises this to 0.998 AUC and 0.951 
P@5. The larger gap between DEMUX-Dir and \textsc{DEMUX} relative to 
that between ARES'25 and ARES'25+BM suggests that BM's benefits are 
amplified when the downstream model is designed to exploit multi-level 
temporal structure. These results support treating BM as a standard 
preprocessing component for any multi-tab WF pipeline that relies on 
temporal or burst-derived features.

\begin{figure}[tb]
  \centering
  \includegraphics[width=\linewidth]{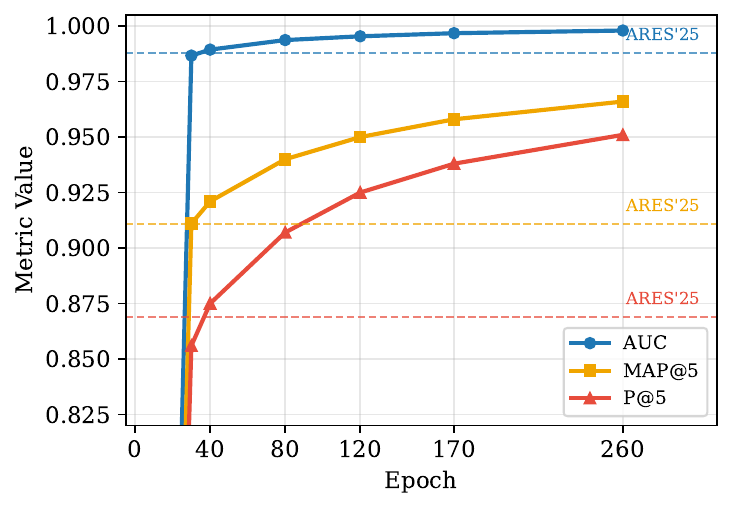}
  \caption{Convergence behaviour of DEMUX in the open-world 5-tab setting.
  The model already exceeds the strongest baseline after 40 epochs and
  saturates near epoch~120.}
  \label{fig:training_curve}
\end{figure}

\subsection{Convergence Analysis}
\label{subsec:convergence}

Figure~\ref{fig:training_curve} shows the learning curve of 
\textsc{DEMUX} in the open-world 5-tab setting. By epoch 40, 
\textsc{DEMUX} already reaches an AUC of 0.989, a P@5 of 0.875, and 
a MAP@5 of 0.921, surpassing ARES'25 across all three metrics before 
training is half complete.

The curve exhibits a clear two-phase shape. In the first phase, 
spanning epochs 0 to 40, all three metrics rise steeply as the local 
modules learn short-range burst patterns from the multi-level input. 
Progress then becomes more gradual as the Transformer-based global 
association module refines long-range dependencies across windows. 
The model approaches saturation near epoch 120 and converges at 
epoch 260 with AUC of 0.998, P@5 of 0.951, and MAP@5 of 0.966.

The two-phase behavior has a practical implication: early stopping 
at epoch 40 already yields performance that surpasses all baselines 
and is suitable for compute-limited deployments, while full training 
to epoch 260 is warranted when maximum precision is required.

\begin{table}[t]
  \centering
  \caption{Feature aggregation in the open-world 5-tab setting.
  Up-proj denotes a learnable pointwise expansion of each position.}
  \label{tab:agg}
  \begin{tabular}{lccc}
    \toprule
    \textbf{Aggregation} & \textbf{AUC} & \textbf{P@5} & \textbf{MAP@5} \\
    \midrule
    Flatten                          & 0.944 & 0.888 & 0.929 \\
    Mean pooling                     & 0.980 & 0.939 & 0.957 \\
    \textbf{Up-proj + mean (DEMUX)}   & \textbf{0.985} & \textbf{0.951} & \textbf{0.966} \\
    \bottomrule
  \end{tabular}
\end{table}

\begin{table}[t]
  \centering
  \caption{Positional encoding in the open-world 5-tab setting.
  RoPE = Rotary Positional Embedding.}
  \label{tab:pos}
  \begin{tabular}{lccc}
    \toprule
    \textbf{Encoding} & \textbf{AUC} & \textbf{P@5} & \textbf{MAP@5} \\
    \midrule
    None                   & 0.996 & 0.937 & 0.956 \\
    Sinusoidal             & 0.996 & 0.936 & 0.955 \\
    Learnable              & 0.996 & 0.934 & 0.954 \\
    \textbf{RoPE (DEMUX)}   & \textbf{0.998} & \textbf{0.951} & \textbf{0.966} \\
    \bottomrule
  \end{tabular}
\end{table}

\subsection{Sensitivity Analysis}
\label{subsec:sensitivity}

We examine two architectural design dimensions under the open-world 
5-tab setting: the strategy used to aggregate the Transformer output 
sequence into a global representation, and the positional encoding 
scheme used in the global association module.

Table~\ref{tab:agg} compares three aggregation strategies. Flattening 
all $L$ position-wise vectors into a single $Ld$-dimensional 
representation preserves positional information explicitly but 
substantially inflates feature dimensionality, yielding an AUC of 
0.944 and a P@5 of 0.888. Mean pooling reduces this to a 
$d$-dimensional descriptor and improves robustness, raising AUC to 
0.980 and P@5 to 0.939, though it treats all positions uniformly and 
may suppress discriminative variation across windows. The proposed 
Up-proj + mean first expands each position-wise feature via a 
learnable pointwise projection before averaging, increasing per-window 
representational capacity while retaining the stability of pooling. 
This yields the best results, with AUC of 0.985 and P@5 of 0.951, 
suggesting that higher-capacity window representations improve 
robustness to the overlapping segmentation and mixed-tab alignment 
that are intrinsic to this setting.

Table~\ref{tab:pos} compares four positional encoding schemes. 
Removing positional encoding entirely remains competitive at AUC of 
0.996 and P@5 of 0.937, indicating that the multi-scale front-end 
already encodes substantial local ordering information. Sinusoidal~\cite{vaswani2017attention} 
and learnable absolute encodings perform slightly worse, with P@5 
dropping to 0.936 and 0.934 respectively, which we attribute to the 
mismatch between fixed absolute position indices and the variability 
introduced by overlapping segmentation and dynamic tab compositions. 
RoPE~\cite{su2024roformer} achieves the best results at AUC of 0.998 
and P@5 of 0.951, as its relative positional offsets align naturally 
with the need to associate burst fragments across overlapping windows 
regardless of their absolute positions in the mixed trace, directly 
addressing (Requirement 3) as motivated in Section~\ref{sec:introduction}.

\subsection{Ablation Analysis}
\label{sec:ablation}

Table~\ref{tab:ablation} reports an ablation study on the open-world 
5-tab setting, isolating the contribution of three architectural 
components: the Boundary Preserving Aggregation Module (BM), the Multi-Scale Parallel CNN (MSP-CNN) in Multi-Granularity Local Analysis, and the Global Association(GA). 
Each variant disables or simplifies exactly one component while keeping 
the others intact.

\begin{table}[t]
  \centering
  \caption{Ablation study in the open-world 5-tab setting.
    \ding{51} = module enabled; \ding{55} = removed or reduced.}
  \label{tab:ablation}
  \resizebox{\columnwidth}{!}{%
  \begin{tabular}{@{}lccccc@{}}
    \toprule
    Variant & BM & MSP-CNN & GA & P@5 & MAP@5 \\
    \midrule
    w/o BM          & \ding{55} & \ding{51} & \ding{51} & 0.876 & 0.907 \\
    kernel=3 only   & \ding{51} & \ding{55} & \ding{51} & 0.924 & 0.946 \\
    kernel=5 only   & \ding{51} & \ding{55} & \ding{51} & 0.926 & 0.949 \\
    kernel=7 only   & \ding{51} & \ding{55} & \ding{51} & 0.919 & 0.943 \\
    w/o Transformer & \ding{51} & \ding{51} & \ding{55} & 0.574 & 0.745 \\
    \textbf{DEMUX (full)} & \ding{51} & \ding{51} & \ding{51} 
                    & \textbf{0.951} & \textbf{0.966} \\
    \bottomrule
  \end{tabular}%
  }
\end{table}

Replacing BM with a plain direction sequence causes P@5 to drop from 
0.951 to 0.876 and MAP@5 from 0.966 to 0.907. This confirms that the 
multi-level representation and overlapping window partitioning provided 
by BM are essential for preserving burst-boundary transition signals 
that fixed non-overlapping segmentation systematically destroys, 
consistent with the (R1) motivation in Section~\ref{sec:introduction}.

Replacing MSP-CNN with a single-kernel branch reduces P@5 to the range 
of 0.919 to 0.926 regardless of which kernel size is retained. The 
similar degradation across all three kernel sizes confirms that no 
single receptive field adequately handles the temporal-scale 
heterogeneity of multi-tab traffic; the parallel multi-scale design is 
necessary to jointly capture fine-grained burst patterns and 
coarse-grained periodic structures, addressing (R2).

Removing the Transformer yields the largest degradation, with P@5 
dropping to 0.574 and MAP@5 to 0.745. Without global association, the 
model cannot link dispersed fingerprint fragments from the same website 
across the mixed trace, confirming that long-range temporal reasoning 
is the most critical capability for multi-tab demixing and that local 
feature extraction alone is fundamentally insufficient.

Taken together, these results confirm that all three components are 
necessary and complementary: BM addresses (R1) by preserving 
boundary-level cues, MSP-CNN addresses (R2) through multi-scale local 
perception, and the Transformer addresses (R3) by integrating long-range 
dependencies across the full sequence.

\section{Discussion}
\label{sec:discussion}

Below, we discuss design choices, limitations, practical considerations, and future research directions.

\textbf{Why a passive-only threat model.}

DEMUX assumes a passive eavesdropper who observes packet metadata without modifying traffic. We adopt this setting because it represents the strictly harder demixing problem. The adversary receives no auxiliary signal beyond the temporally superimposed mixture of flows, so the entire burden of traffic demixing falls on the model architecture rather than on active probing. Our results demonstrate that this burden is well handled by the proposed co-design. Even in the most challenging 5-tab closed-world setting, DEMUX attains a P@5 of 0.943 and MAP@5 of 0.961 under purely passive observation. Under TrafficSliver, where the adversary observes only a partial fraction of packets at any single entry node, it still maintains a P@2 of 0.940. These results suggest that the pipeline of the Boundary Preserving Aggregation Module, the Multi-Scale Parallel CNN, and the RoPE-enhanced Transformer encoder captures sufficient discriminative structure in multi-tab traffic to achieve effective demixing without relying on active traffic manipulation, which would risk detection by Tor's integrity checks or network anomaly monitors. An active adversary could further inject probing packets or manipulate timing to facilitate demixing. Incorporating such signals is a natural extension but lies outside the structural contributions demonstrated here.

\textbf{Implications of the demixing perspective.}

Treating multi-tab WF as an implicit traffic demixing problem changes the view of the task. Rather than a harder version of single-tab classification, the problem becomes closer in structure to blind source separation in signal processing. This view yields two implications that extend beyond the specific architecture of DEMUX. First, it helps explain why prior single-tab-derived architectures plateau as tab count grows. The absence of boundary-aware aggregation and relative-position reasoning is not a tuning issue but a structural mismatch with the demixing task. Second, it suggests that future WF research may benefit from treating temporal mixing as a first-class design constraint rather than as noise to be absorbed by model capacity. We view the three structural requirements identified in this work (R1 to R3) as a starting set, and we expect further requirements to emerge as the community explores richer observation modalities.

\section{Conclusion}
\label{sec:conclusion}

This paper argues that multi-tab website fingerprinting is not a harder instance of single-tab classification but a qualitatively different problem. It is a problem of implicit traffic demixing, in which the adversary must recover an unknown number of source identities from a single superimposed observation with no explicit boundary cues. Posing the task in these terms exposes three structural requirements that effective architectures need to jointly satisfy, and reveals that no prior method does so. DEMUX is the first framework designed from this perspective rather than inherited from the single-tab lineage, and each of its components, namely overlapping boundary-preserving aggregation, multi-scale parallel convolution, and a two-stage RoPE-enhanced Transformer encoder, is derived directly from one of these requirements.
Empirically, DEMUX establishes a new state of the art across closed-world, open-world, defense-augmented, dynamic-tab, and cross-configuration settings. More importantly, it exhibits the slowest performance degradation as mixing intensifies, indicating that its advantage stems from structural alignment with the demixing task rather than from incremental capacity gains. A further practical contribution is that the Boundary Preserving Aggregation Module transfers cleanly to existing architectures including DF, TMWF, and ARES'25, delivering consistent improvements without any modification to the downstream model and serving as a reusable preprocessing baseline for the field.
Beyond the specific results, we hope the demixing perspective invites a broader rethinking of how multi-tab traffic analysis should be posed, evaluated, and defended against. The three structural requirements identified here are unlikely to be exhaustive, and we expect that richer observation modalities, longer sessions, and stronger defenses will motivate additional requirements in future work. Treating multi-tab traffic as a demixing problem, rather than as a noisy multi-label classification problem, provides what we believe is a more faithful foundation on which the next generation of website fingerprinting research can build.

\section*{Acknowledgment}
This work was supported in part by the National Key Research and Development Program of China (Grant No. 2023YFB3106700) under the Young Scientists Program, in part by Natural Science Foundation of Jiangsu Province (Grant No. SBK2023041256), in part by the National Natural Science Foundation of China (Grant No. 62302097), and in part by the National Undergraduate Training Programs for Innovation.

\bibliographystyle{IEEEtran}
\bibliography{ref_list}

\vfill

\end{document}